\pdfoutput=1

\documentclass[11pt]{article}

\usepackage[]{ACL2023}

\usepackage{times}
\usepackage{latexsym}
\usepackage{amsmath}
\usepackage{amssymb}
\usepackage{booktabs,tabularx}
\usepackage{graphicx}
\usepackage{tikz}
\usepackage{pifont}
\usepackage{multirow}
\usepackage{pdfpages}
\usepackage{listings}
\usepackage{xcolor, colortbl}

\usepackage{tcolorbox}
\tcbuselibrary{listings,skins}
\tcbuselibrary{skins, listings, breakable}
\newtcolorbox{promptbox}[2][]{%
  float*=htb,
  floatplacement=htb,
  width=\textwidth,
  before skip=1em, after skip=1em,
  enhanced, breakable,
  colback=gray!5!white, 
  colframe=gray!75!black,
  colbacktitle=gray!75!black, 
  coltitle=white, 
  fonttitle=\bfseries,
  title={#2},      
  label={#1},      
  listing engine=listings,  
  listing only,
  listing options={
    basicstyle=\ttfamily\small,
    breaklines=true,        
    breakanywhere=true,     
  },
}

\usepackage{subcaption}

\usetikzlibrary{positioning, shapes.geometric, arrows.meta}

\usepackage[T1]{fontenc}

\usepackage[utf8]{inputenc}

\usepackage{microtype}

\usepackage{inconsolata}

\title{DocMMIR: 
A Framework for Document Multi-modal Information Retrieval}

\newcommand*\samethanks[1][\value{footnote}]{\footnotemark[#1]}
\author{
\textbf{Zirui Li}\textsuperscript{1}\Thanks{ Equal Contribution.}\quad 
\textbf{Siwei Wu}\textsuperscript{1}\samethanks[1]\quad 
\textbf{Yizhi Li}\textsuperscript{1}\samethanks[1]\quad 
\textbf{Xingyu Wang}\textsuperscript{2} \quad 
\textbf{Yi Zhou}\textsuperscript{2} \quad 
\textbf{Chenghua Lin}\textsuperscript{1}\Thanks{ Corresponding Author.}\quad 
\\
\textsuperscript{1}University of Manchester\quad
\textsuperscript{2}University of Cardiff\quad 
\\
\texttt{zirui\_li\_0806@outlook.com}
\\
\texttt{\{siwei.wu-2, chenghua.lin\}@manchester.ac.uk}
}

\begin{document}
\maketitle
\begin{abstract}
The rapid advancement of unsupervised representation learning and large-scale pre-trained vision-language models has significantly improved cross-modal retrieval tasks. However, existing multi-modal information retrieval (MMIR) studies lack a comprehensive exploration of document-level retrieval and suffer from the absence of cross-domain datasets at this granularity. To address this limitation, we introduce \textbf{DocMMIR}, a novel multi-modal document retrieval framework designed explicitly to unify diverse document formats and domains—including Wikipedia articles, scientific papers (arXiv), and presentation slides—within a comprehensive retrieval scenario. We construct a large-scale cross-domain multi-modal benchmark, comprising \textbf{450K} samples which systematically integrates textual and visual information. Our comprehensive experimental analysis reveals substantial limitations in current state-of-the-art MLLMs (CLIP, BLIP2, SigLIP-2, ALIGN) when applied to our tasks, with only CLIP demonstrating reasonable zero-shot performance.  
Furthermore, we conduct a systematic investigation of training strategies, including cross-modal fusion methods and loss functions, and develop a tailored approach to train CLIP on our benchmark. This results in a \textbf{+31\%} improvement in MRR@10 compared to the zero-shot baseline. All our data and code are released in \url{https://github.com/J1mL1/DocMMIR}.

\end{abstract}

\begin{figure*}[ht]
\centering
\includegraphics[width=0.9\textwidth]{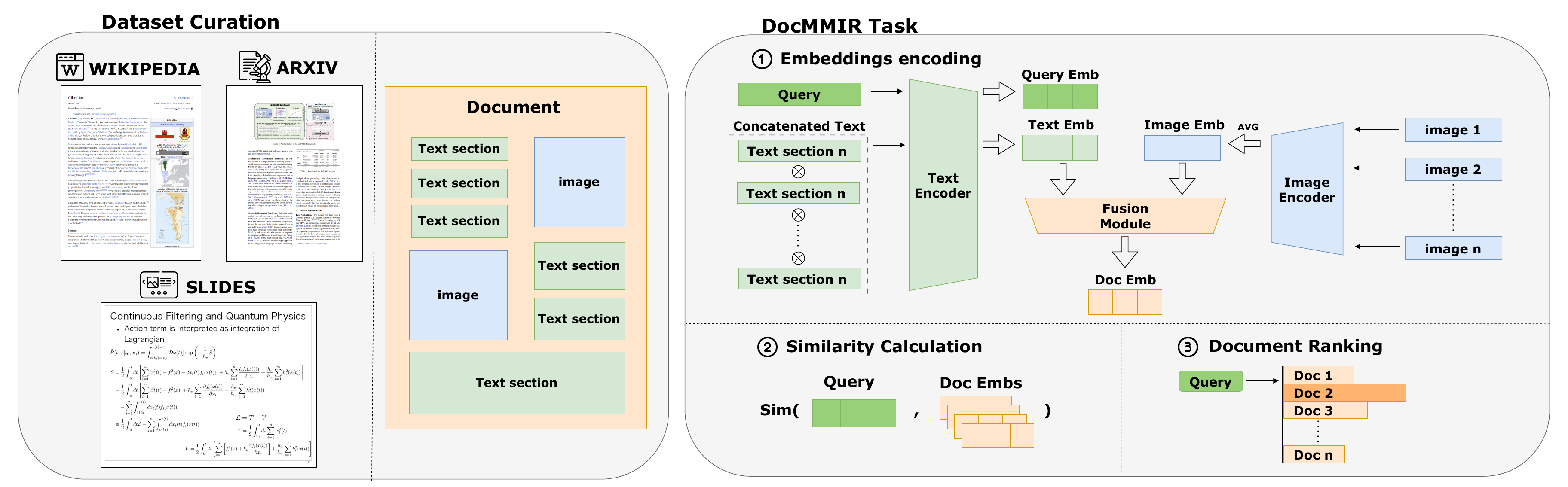}
\caption{Overview of DocMMIR. The multi-modal documents across domains are formalized into a unified framework.}
\label{fig:overview}
\vspace{-4mm}
\end{figure*}

\section{Introduction}

With the emergence of the Large Pre-training Model, the Pre-trained Multi-modal Retrieval Models (i.e.,  CLIP \cite{clip}) and Large Visual Language Models (i.e., BLIP2 \cite{blip2}, ALIGN \cite{align}), trained on vast image-text corpora, have achieved state-of-the-art results on tasks such as cross-modal retrieval, image captioning, and visual question answering by capturing fine-grained cross-modal alignment.

Despite these advances in image–text retrieval, applying MLLMs to conduct document-level retrieval remains challenging. Real-world documents (Wikipedia articles, scientific papers, slide decks, etc.) often contain complex page layouts with text, figures, tables, and other formats. Off-the-shelf MLLMs typically encode images or text snippets in isolation and lack awareness of longer document context or multi-page structure. Existing studies fail to fully address the complexity of real-world document structures, leading to several critical limitations, as illustrated in \autoref{tab:mmir_benchmarks}. First, most multi-modal document retrieval benchmarks are constrained to fine-grained snippet-level settings, lacking support for whole-document retrieval. Second, the diversity of domains represented in current datasets is limited—many focus on narrow scientific or web-centric sources, overlooking broader cross-domain applicability. Third, some approaches treat whole-document screenshots as unified inputs and show promising results on specific domains, but this strategy assumes visual capture and still ignores semantic layout. These limitations hinder the comprehensive evaluation and development of MLLMs for document-level multi-modal information retrieval.

To address these gaps, we introduce \textbf{DocMMIR}: a unified framework for \textbf{Doc}ument-level \textbf{M}ulti-\textbf{M}odal \textbf{I}nformation \textbf{R}etrieval across various domains (demonstrated in \autoref{fig:overview}). We curate a large-scale benchmark with a unified structure that includes documents and queries from diverse sources (Wikipedia articles, arXiv papers, and slide presentations), each containing a mixture of modalities. Our retrieval model extends a dual-encoder MLLM's architecture to operate on entire documents: during training, we optimise a contrastive training objective over query–document pairs by fusing embeddings from images and texts to construct doc embeddings, which are meant to distill information and be a unified representation. This unified model is trained end-to-end on the DocMMIR dataset spanning all domains and modalities. Experimental results show that fine-tuning models on DocMMIR improves retrieval performance over corresponding zero-shot baselines. In particular, fine-tuning the CLIP encoder on our document-level tasks yields large gains in MRR@10 across domains. We also find that incorporating multi-modal content is critical: purely text-based retrieval falls far short of our multi-modal model, and naive image-only search misses context. Our contributions are thus threefold:

\noindent\textbf{Framework.} We design a unified modeling framework for multi-modal retrieval that can accommodate and jointly encode diverse document formats, including academic papers, presentation slides, and Wikipedia pages, by leveraging both textual content and associated imagery in each document.

\noindent\textbf{Dataset.} A large-scale cross-domain multi-modal dataset is constructed, combining multiple sources of multi-modal documents into a single training and evaluation suite. This dataset enables consistent evaluation of retrieval methods across domains and provides a foundation for training models to be generally applicable. 

\noindent\textbf{Modeling.} Various unified modeling strategies have been explored, including multi-modal fusion strategies and loss function selection, to understand what training paradigms yield the best retrieval performance for this task.

\begin{table*}[htb]
\centering
\resizebox{\linewidth}{!}{%
\begin{tabular}{l c l l c l l}
\toprule
\textbf{Benchmark} &
\textbf{Num of Query (Train/Val/Test)} &
\textbf{Tasks} &
\textbf{Domain} &
\textbf{Doc-level} &
\textbf{Annot.} &
\textbf{Query} \\
\midrule
DocCVQA \cite{doccvqa} & -- / -- / 20 & Doc VQA & Finance docs & \textcolor{red}{\ding{55}} & Expert & Human \\
SciMMIR \cite{scimmir} & 498.3K / 16.4K / 16.3K & Fig–caption retrieval & Scientific papers & \textcolor{red}{\ding{55}} & Automatic & Synthetic \\
ColPali \cite{colpali} & 127.5K / -- / 2.8K & Doc retrieval & Multi-domain & \textcolor{green}{\ding{51}} & Both & Both \\
PDF-MVQA \cite{mvqa} & 180.8K / 27.6K / 54.5K & Doc VQA & Biomedical PDFs & \textcolor{red}{\ding{55}} & Automatic & Synthetic \\
MMLongBench-Doc \cite{mmlongbench} & -- / -- / 1.1K & Doc VQA & Multi-domain & \textcolor{red}{\ding{55}} & Expert & Human \\
MMDocIR \cite{mmdocir} & 73.8K / -- / 1.6K & Page/Layout retrieval & Multi-domain & \textcolor{green}{\ding{51}} & Both & Both \\
DSE \cite{dse} & 10K / -- / 5.7K & Doc retrieval & Wiki + Slides & \textcolor{green}{\ding{51}} & Automatic & Synthetic \\
UniIR (M-BEIR) \cite{uniir} & 1.1M / 182K / 190K & Multi-task MMIRs & Multi-domain & \textcolor{red}{\ding{55}} & Automatic & Synthetic \\
VLM2Vec (MMEB) \cite{vlm2vec} & -- /-- /--  & Multi-task MMIRs & Multi-domain & \textcolor{red}{\ding{55}} & Automatic & Synthetic \\
MARVEL (ClueWeb22-MM) \cite{marvel} & 72.0K / 10K / 10K & Fig-caption retrieval & Web pages & \textcolor{red}{\ding{55}} & Automatic & Synthetic \\
IDentIfy \cite{unified} & 390.5K / 26.8K / 79.4K & Doc retrieval & Wiki & \textcolor{green}{\ding{51}} & Automatic & Synthetic \\
MM-Embed \cite{mmbed} & — / — / — & Multi-task MMIRs & Multi-domain & \textcolor{red}{\ding{55}} & Automatic & Synthetic \\
GME \cite{gme} & — / — / 200K & Multi-task MMIRs & Multi-domain & \textcolor{red}{\ding{55}} & Automatic & Synthetic \\
\midrule
DocMMIR (Ours) & 450.1K / 19.2K / 19.2K & Doc retrieval & Wiki/ArXiv/Slides & \textcolor{green}{\ding{51}} & Automatic & Synthetic \\
\bottomrule
\end{tabular}}
\caption{Consolidated multi-modal IR benchmarks.}
\label{tab:mmir_benchmarks}
\vspace{-5mm}
\end{table*}

\vspace{-3mm}

\section{Task Definition}
The DocMMIR task aims to retrieve semantically relevant documents from a heterogeneous collection spanning Wikipedia articles, arXiv papers, and presentation slides—each combining text (e.g., paragraphs, bullet points, section headers) with visual elements (e.g., figures, diagrams). Given a text query generated by a large language model and paired with a document, the system must rank all documents and identify the correct one. These domains differ in structure and modality: arXiv emphasizes technical terms and plots, slides prioritize visuals and brevity, while Wikipedia blends narrative text with images. This task poses several challenges: (1) cross-domain heterogeneity in layout, density, and style; (2) modality imbalance between text- and image-heavy documents; and (3) fine-grained alignment, where queries may refer to specific figures or terms, requiring joint modeling of both modalities.

Formally, let $\mathcal{D} = \{D_1, D_2, \dots, D_N\}$ denote a cross-domain document collection, where each document $D_k$ comprises a set of text segments $\{t_i\}_{i=1}^N$ and images $\{i_j\}_{j=1}^M$. A query $q$, generated to holistically describe the document's content (e.g., \textit{``Retrieve a paper explaining transformer-based vision models with ablation studies''}), serves as the input. 
The retrieval system computes a relevance score $\text{sim}(q, D_k)$ for each document, producing a ranked list $\mathcal{R} = [D_{(1)}, D_{(2)}, \dots, D_{(N)}]$ where higher-ranked documents better satisfy the intent.

\section{Data Curation Pipeline}
To enable heterogeneous sources to be standardized into a unified representation, we introduce a domain-specific but modular preprocessing workflow for different data sources.
During the data processing phase, we design a multi-modal quality control mechanism combining tokeniser-aware statistics, perplexity scores, and garbledness heuristics, particularly beneficial for noisy OCR content. 
We also incorporate a human-in-the-loop feedback for the retrieval query generation to raise the standard of annotation. 
More importantly, our pipeline is extendable: it can be directly adapted to new domains by plugging in domain-specific parsing and applying the same unification logic. 
\vspace{-3mm}

\subsection{Overview and Data Sources}

To support document-level multi-modal retrieval, we curate a large-scale dataset composed of Wikipedia articles, arXiv scientific papers, and Slideshare presentations. These three domains span a wide spectrum of content styles and multi-modal characteristics, providing a robust foundation for cross-domain evaluation. The Wikipedia portion is derived from the WIT dataset~\cite{Srinivasan2021WIT}, from which we retain English entries with at least one image and more than 300 tokens to ensure sufficient multi-modal context. The arXiv data comes from the SciMMIR dataset~\cite{scimmir}, consisting of figure-caption-text triples extracted from scientific documents with structured formatting and academic diagrams. For Slideshare, we use the public Slideshare-1M dataset~\cite{Chen2016Slideshare}, containing over 30,000 slide decks. We apply OCR to extract text from each slide and pair it with its corresponding image, transforming slide sequences into coherent multi-modal documents.

\subsection{Domain-specific Preprocessing Strategies}

Each domain presents unique structural and quality challenges, requiring targeted preprocessing steps to ensure coherence, multi-modal richness, and compatibility with downstream tasks.

\noindent\textbf{Wikipedia} articles often contain sparse images and loosely connected paragraphs. To construct coherent multi-modal documents, we filter for articles with at least one image and over 300 tokens. We then assemble all paragraphs and images under a shared title to strengthen semantic alignment between text and image. This preserves entity-rich content and factual consistency, making Wikipedia ideal for knowledge-grounded retrieval.

\noindent\textbf{arXiv} papers offer well-structured figures and technical writing but include noisy elements such as LaTeX math. We remove inline and block-level math expressions (e.g., \verb|$...$|, \verb|\begin{equation}|) to reduce tokenisation noise. Paragraphs are retained only if they co-occur with figures in the same section, forming meaningful figure-text pairs. This allows us to build documents that capture the scientific logic and visual reasoning of academic writing.

\noindent\textbf{Slideshare} presentations consist of fragmented slides with dense imagery and varied textual quality. We apply OCR to extract text, then use perplexity-based filtering (via Qwen3 \cite{qwen3}) and garbled content detection to remove noisy slides. Specifically, we detect LaTeX artifacts and substrings with excessive special characters (\verb|[^\w\s]{10,}|), removing documents where over 50\% of paragraphs are garbled. Finally, slides from each deck are flattened into single document-level units to enhance semantic continuity. This conversion from slide-level to document-level structure is essential for retrieval tasks.

\subsection{Unified Representation and Quality Filtering}

To standardize the dataset across domains, we adopt a unified document format consisting of: (1) a sequence of text blocks tokenised using the \texttt{cl100k\_base} tokeniser; (2) a list of aligned images (all non-empty); and (3) one or more natural-language queries per document. Text statistics such as paragraph length, special character ratio, and query length are computed uniformly using token-level counts. This unification enables all documents, regardless of domain, to be processed by the same retrieval architecture.

We enforce consistent quality control across domains. Documents with insufficient image-text pairs, excessive garbled content, or high perplexity segments are discarded. Only fully multi-modal, semantically coherent documents are retained, ensuring that the resulting dataset supports robust, generalizable retrieval training.

\subsection{Dataset Statistics and Distribution}

The final dataset contains 488,467 documents, split into 450,079 for training, 19,184 for validation, and 19,204 for testing. Wikipedia contributes the largest portion (389,838 documents), characterized by moderate text length (380 tokens) and low image density (1.33 images/document). arXiv provides 68,764 documents, with an average of 7.72 images and 765.96 tokens per document, reflecting the formatting of scientific publications. Slideshare contributes 29,865 documents, notable for their high image density (30.43 images/document) and long OCR-derived text (2,060.42 tokens/document). Across all domains, query lengths range from 39.35 to 54.70 tokens, ensuring a consistent information-seeking format for retrieval tasks.

\subsection{Query Generation and Annotation}

Each document is annotated with one or more natural-language queries generated using Qwen2.5-VL \cite{qwen2.5,qwen2} under a unified prompt template (Appendix~\ref{app:prompt}). To ensure high-quality supervision, 2,000 sampled queries were reviewed by three NLP-trained annotators. A query is marked “highly relevant” if positively rated by at least two reviewers. Overall, 87.5\% of queries met this standard. Low-quality queries were revised and used to refine future prompts, forming a feedback loop. We encourage queries to be paraphrased and abstract to increase retrieval difficulty and realism.

\begin{table}[ht]
  \centering
  \label{tab:full-stats-vertical}
  \resizebox{\linewidth}{!}{%
  \begin{tabular}{lcccc}
    \toprule
    \textbf{Statistic} & \textbf{Wiki} & \textbf{ArXiv} & \textbf{Slide} & \textbf{Total} \\
    \midrule
    \#Train       & 360{,}285 & 62{,}764 & 27{,}057 & 450{,}079 \\
    \#Valid       & 14{,}775  & 3{,}000  & 1{,}409  & 19{,}184  \\
    \#Test        & 14{,}805  & 3{,}000  & 1{,}399  & 19{,}204  \\
    \#Total Doc        & 389{,}838  & 68{,}764  & 29{,}865  & 488{,}467  \\    
    Avg.\#Images  & 1.33      & 7.72     & 30.43    & 4.01        \\
    Avg.\#Texts   & 380.44    & 765.96   & 2060.42  & 537.43        \\
    Avg.\#Query   & 43.71     & 54.70    & 39.35    & 44.99        \\
    \bottomrule
  \end{tabular}}
  \caption{Comprehensive token-level statistics of the DocMMIR dataset by domain. All token counts are computed using the \texttt{cl100k\_base} tokeniser.}
\vspace{-5mm}
\end{table}

\section{Methodology}
\label{methodology}
\textbf{DocMMIR} is designed to perform document-level multi-modal information retrieval by jointly leveraging textual and visual information. It adopts a dual-encoder architecture with late fusion and is trained using a symmetric batch-wise binary classification objective.

\paragraph{Modality Encoding.}
Each document consists of textual segments \(\{t_1, \dots, t_N\}\) and image elements \(\{i_1, \dots, i_M\}\). The textual content is concatenated and encoded using a pretrained transformer encoder, from which the \texttt{[CLS]} token is extracted as the document-level text representation:
\begin{equation}
E_{\text{text}} = \text{Encoder}_{\text{text}}\left(\bigoplus_{i=1}^N t_i\right)
\label{eq:text_embedding}
\end{equation}
Each image is encoded independently, and the final image representation is computed as the mean of all image embeddings:
\begin{equation}
E_{\text{img}} = \frac{1}{M} \sum_{j=1}^M \text{Encoder}_{\text{img}}(i_j)
\label{eq:image_embedding}
\end{equation}

Similarly, each query \(q\) is a short textual input associated with a document and is encoded using the same text encoder. The final query embedding is derived from its \texttt{[CLS]} token:
\begin{equation}
E_{\text{query}} = \text{Encoder}_{\text{text}}(q)
\label{eq:query_embedding}
\end{equation}
The same encoder is used for both document text and query encoding to ensure consistent representation space.

\paragraph{Fusion Strategy.}
To obtain a unified document representation, we apply a weighted-sum fusion of the text and image embeddings:
\begin{equation}
E_{\text{doc}} = \alpha \cdot E_{\text{text}} + (1 - \alpha) \cdot E_{\text{img}}
\label{eq:weighted_sum}
\end{equation}
where \(\alpha \in [0, 1]\) is a fixed hyperparameter controlling the relative contribution of each modality. This late fusion strategy is simple and effective, and offers robust performance across domains with diverse modality densities.

\paragraph{Training Objective.}
The model is trained using a symmetric batch-wise binary cross-entropy (BCE) loss, which treats each query–document pair as a binary classification task. For a batch of size \( B \), let \( S \in \mathbb{R}^{B \times B} \) denote the cosine similarity matrix between all query and document embeddings. Let \( T = I_B \) be the identity matrix marking ground-truth positive pairs. The loss is computed as:

\begin{equation}
\resizebox{0.95\columnwidth}{!}{$
  \mathcal{L}_{\mathrm{BCE}}
  = \frac{1}{2}\Bigl[
    \mathrm{BCE}\bigl(\sigma(S), T\bigr)
    + \mathrm{BCE}\bigl(\sigma(S^\top), T\bigr)
  \Bigr]
$}
\label{eq:bce_loss}
\end{equation}

where \( \sigma(\cdot) \) denotes the element-wise sigmoid function. This symmetric formulation ensures that both the query-to-document and document-to-query directions are equally optimized.

The binary cross-entropy for each element \( s_{ij} \in S \) and label \( t_{ij} \in T \) is defined as:

\begin{equation}
\resizebox{0.95\columnwidth}{!}{$
\text{BCE}(s_{ij}, t_{ij}) = - \left[
t_{ij} \cdot \log(s_{ij}) + (1 - t_{ij}) \cdot \log(1 - s_{ij})
\right],
$}
\end{equation}
This objective encourages the model to assign high similarity scores to matching query–document pairs (\(t_{ij} = 1\)) and low scores to all other combinations (\(t_{ij} = 0\)). The loss is averaged over all \( B^2 \) elements in the similarity matrix.

During inference, retrieval is performed by computing cosine similarity between the encoded query and document embeddings:
\[
\mathrm{sim}(q, d) = \frac{E_{\text{query}} \cdot E_{\text{doc}}}{\|E_{\text{query}}\| \cdot \|E_{\text{doc}}\|}.
\]
Finally, Documents are ranked by similarity scores to produce the final results.

\begin{table*}[htb]
\centering
\resizebox{0.9\linewidth}{!}{%
\begin{tabular}{l ccc cccc}
\toprule
\multirow{2}{*}{\textbf{Model}} 
  & \textbf{Wiki} & \textbf{ArXiv} & \textbf{Slides}
  & \multicolumn{4}{c}{\textbf{Fullset}} \\
\cmidrule(lr){2-4} \cmidrule(lr){5-8}
  & \multicolumn{3}{c}{\textbf{MRR@10}} 
  & \textbf{HIT@1} & \textbf{HIT@3} & \textbf{MRR@10} & \textbf{NDCG@10} \\
\midrule
CLIP (ViT-B/32) \cite{clip}
  & 0.00280 & 0.00163 & 0.00062 
  & 0 & 0.00031 & 0.00025 & 0.00041 \\

CLIP (ViT-L/14) \cite{clip}
  & 0.32467 & 0.09516 & 0.19584 
  & 0.20133 & 0.31089 & 0.27039 & 0.30911 \\

ALIGN \cite{align}
  & 0.00192 & 0.00246 & 0.00053 
  & 0.00062 & 0.00125 & 0.00108 & 0.00145 \\

SigLIP-2 \cite{siglip2}
  & 0.00621 & 0.00437 & 0.00220 
  & 0.00099 & 0.00187 & 0.00176 & 0.00231 \\

BLIP-2 \cite{blip2}
  &    0   &    0.00098   &    0.00036   
  &    0.00002   &    0   &    0.00005   &   0.00004   \\

\midrule
\textbf{FT CLIP (ViT-L/14)} 
  & 0.77280 & 0.40380 & 0.32230 
  & 0.60789 & 0.71444 & 0.66945 & 0.69929 \\
\bottomrule
\end{tabular}}
\caption{Overall DocMMIR benchmark results. The first header row lists the three test domains and the fullset; the second row specifies that we report MRR@10 for each domain and four metrics for the fullset.}
\label{tab:docmmir_overall}
\vspace{-5mm}
\end{table*}

\begin{table*}[htb]
\centering
\resizebox{0.7\linewidth}{!}{%
\begin{tabular}{llccccc}
\toprule
\textbf{Method} & \textbf{Dataset} & \textbf{MRR@10} & \textbf{HIT@1} & \textbf{HIT@3} & \textbf{HIT@10} & \textbf{NDCG@10} \\
\midrule
\multirow{4}{*}{Weighted Sum + BCE} 
  & Full     & 0.65662 & 0.60022 & 0.69651 & 0.77123 & 0.68423 \\
  & Wiki     & 0.75703 & 0.70541 & 0.79773 & 0.85314 & 0.78042 \\
  & ArXiv    & 0.39097 & 0.31239 & 0.43140 & 0.57612 & 0.43495 \\
  & Slide    & 0.30944 & 0.24895 & 0.34227 & 0.45575 & 0.34422 \\
\midrule
\multirow{4}{*}{Weighted Sum + InfoNCE} 
  & Full     & 0.63606 & 0.57523 & 0.67888 & 0.76077 & 0.66614 \\
  & Wiki     & 0.74367 & 0.68775 & 0.78673 & 0.85147 & 0.76984 \\
  & ArXiv    & 0.38550 & 0.29655 & 0.44017 & 0.58378 & 0.43297 \\
  & Slide    & 0.26701 & 0.19672 & 0.30689 & 0.42667 & 0.30510 \\
\midrule
\multirow{4}{*}{MLP + BCE} 
  & Full     & 0.23305 & 0.15745 & 0.25536 & 0.38056 & 0.25932 \\
  & Wiki     & 0.27166 & 0.19606 & 0.31259 & 0.45601 & 0.31533 \\
  & ArXiv    & 0.07154 & 0.03833 & 0.08455 & 0.16268 & 0.09292 \\
  & Slide    & 0.10521 & 0.06818 & 0.12059 & 0.20807 & 0.12930 \\
\midrule
\multirow{4}{*}{MLP + InfoNCE} 
  & Full     & 0.20686 & 0.14283 & 0.24069 & 0.36595 & 0.24453 \\
  & Wiki     & 0.24815 & 0.17407 & 0.28875 & 0.42880 & 0.29100 \\
  & ArXiv    & 0.09621 & 0.05860 & 0.11193 & 0.20102 & 0.12077 \\
  & Slide    & 0.10925 & 0.06944 & 0.12582 & 0.21374 & 0.13381 \\
\bottomrule
\end{tabular}}
\caption{Comparison of fusion and loss strategies on the DocMMIR benchmark.}
\label{tab:fusion_loss_comparison}
\vspace{-5mm}
\end{table*}

\section{Experiment Setting}
\subsection{Baseline Models}
We evaluate DocMMIR against state-of-the-art vision-language models in a zero-shot retrieval setting, including CLIP \cite{clip}, BLIP-2 \cite{blip2}, SigLIP-2 \cite{siglip2} and ALIGN \cite{align}. These baselines span diverse architectures: CLIP and ALIGN rely on contrastive image-text pretraining, BLIP-2 integrates the Qformer to tune vision encoders with language models, and SigLIP-2 introduces sigmoid loss for improved multi-modal alignment. For fairness, all models utilize publicly released checkpoints without domain-specific adaptation.
\subsection{Fine-tuning Strategy}
We fine-tune OpenCLIP \cite{openclip} with the ViT-L/14 backbone, which demonstrated superior zero-shot capabilities on our preliminary benchmarks. The model is initialized from the LAION-2B pretrained weights and adapted to the DocMMIR task via dual-encoder fine-tuning: the text encoder processes concatenated document segments, while the image encoder handles averaged visual features as illustrated in 
\autoref{methodology}.
\subsection{Training configuration}
The model optimization employed AdamW with modality-specific learning rates: 2e-5 for both image and text encoders, enabling balanced parameter updates across visual and textual streams. We applied \(L2\) regularization through a weight decay of 0.01 and stabilized numerical operations with epsilon set to 1e-8. The training schedule incorporated linear learning rate warmup during the initial 10\% of optimization steps, followed by cosine decay. Various batch sizes were implemented to maximize GPU memory utilization on our 8×H100 setup, with automatic mixed precision (AMP) accelerating training while maintaining stability. All runs were limited to 5 epochs maximum duration with early stopping triggered after 2 epochs of validation MRR degradation, ensuring efficient resource use while preventing overfitting.

\subsection{Evaluation Metrics}
Retrieval effectiveness is reported with MRR@10, NDCG@10 and HIT@{\{}1, 3, 10{\}}.  MRR@10 serves as our principal indicator because it reflects user behavior on the first results page, while NDCG@10 captures graded relevance and HIT@K offers precision at exact cut-offs.  A complete description of the evaluation protocol and metric computation is provided in Appendix~\ref{app:eval}.

\begin{table*}[htb]
\centering
\resizebox{0.8\linewidth}{!}{%
\begin{tabular}{llr|cccccc}
\toprule
\textbf{Train Domain} & \textbf{Data \%} & \textbf{Train Size} 
& \textbf{arXiv (3K)} & \textbf{Slides (1.5K)} & \textbf{Wiki (15K)} & \textbf{All (17.5K)} & \textbf{Weighted Avg.} \\
\midrule
\multirow{5}{*}{ArXiv} 
  & 20  & 12,540  & 0.7236 & 0.5810 & 0.5766 & 0.5610 & 0.6681 \\
  & 40  & 25,080  & 0.7439 & 0.5850 & 0.5360 & 0.5377 & 0.6371 \\
  & 60  & 37,620  & 0.7568 & 0.5765 & 0.5824 & 0.5876 & 0.6784 \\
  & 80  & 50,160  & 0.7560 & 0.5691 & 0.5373 & 0.5426 & 0.6389 \\
  & 100 & 62,700  & \cellcolor{yellow!25}0.7644 & 0.5591 & 0.6194 & 0.6176 & 0.7099 \\
\midrule
\multirow{5}{*}{Slides} 
  & 20  & 5,600   & 0.4741 & 0.6794 & 0.6495 & 0.6029 & 0.6962 \\
  & 40  & 11,200  & 0.4545 & 0.6868 & 0.6364 & 0.5930 & 0.6823 \\
  & 60  & 16,800  & 0.4222 & 0.6922 & 0.6075 & 0.5673 & 0.6524 \\
  & 80  & 22,400  & 0.4120 & \cellcolor{yellow!25}0.7064 & 0.5950 & 0.5535 & 0.6412 \\
  & 100 & 28,000  & 0.4036 & 0.6857 & 0.6234 & 0.5709 & 0.6623 \\
\midrule
\multirow{5}{*}{Wikipedia} 
  & 20  & 70,000   & 0.5471 & 0.5273 & 0.8823 & \cellcolor{yellow!25}0.7944 & \cellcolor{yellow!25}0.8952 \\
  & 40  & 140,000  & 0.5065 & 0.5158 & 0.8792 & 0.7843 & 0.8846 \\
  & 60  & 210,000  & 0.5187 & 0.5170 & 0.8800 & 0.7872 & 0.8875 \\
  & 80  & 280,000  & 0.4998 & 0.5145 & \cellcolor{yellow!25}0.8844 & 0.7875 & 0.8878 \\
  & 100 & 350,000  & 0.4487 & 0.4937 & 0.8826 & 0.7762 & 0.8758 \\
\bottomrule
\end{tabular}}
\caption{Effect of training domain and scale across test domains. Best scores per column are highlighted.}
\label{tab:training_scale_hit10}
\vspace{-5mm}
\end{table*}

\begin{table}[ht]
\centering
\resizebox{\linewidth}{!}{%
\begin{tabular}{llcccc}
\toprule
\textbf{Training Set} & \textbf{Modality} & \textbf{Wiki} & \textbf{ArXiv} & \textbf{Slides} & \textbf{Full Set} \\
\midrule
\multirow{2}{*}{Wiki}    
  & Text Only   & \cellcolor{yellow!25}0.63850 & 0.16370 & 0.35085 & 0.53140 \\
  & Image Only  & 0.22230 & 0.06210 & 0.25230 & 0.18990 \\
\midrule
\multirow{2}{*}{ArXiv}    
  & Text Only   & 0.48360 & \cellcolor{yellow!25}0.38550 & 0.38850 & 0.43210 \\
  & Image Only  & 0.27480 & 0.17175 & 0.25590 & 0.24590 \\
\midrule
\multirow{2}{*}{Slides}    
  & Text Only   & 0.43280 & 0.23930 & \cellcolor{yellow!25}0.46610 & 0.39010 \\
  & Image Only  & 0.35550 & 0.17007 & 0.38570 & 0.31700 \\
\midrule
\multirow{2}{*}{Full Set} 
  & Text Only   & 0.62860 & 0.24140 & 0.32970 & \cellcolor{yellow!25}0.53330 \\
  & Image Only  & 0.14090 & 0.03784 & 0.18625 & 0.12290 \\
\bottomrule
\end{tabular}}
\caption{Effect of training set and modality on \textbf{MRR@10} performance across test domains. Best scores per column are highlighted.}
\label{tab:mrr10_modality_ablation}
\vspace{-8mm}
\end{table}

\section{Result Analysis} 

\subsection{Overall Performance Benchmark}

Zero-shot evaluations of pre-trained vision-language models reveal a significant performance gap in the DocMMIR task. As shown in \autoref{tab:docmmir_overall}, most models—such as CLIP (ViT-B/32), ALIGN, and SigLIP-2—achieve near-zero MRR@10 and HIT@K scores across all domains. CLIP (ViT-L/14) is a partial exception, performing modestly on Wikipedia (MRR@10: 0.3247), likely due to its text-focused training. However, it still struggles with domain shifts, dropping to 0.1958 on Slides. In contrast, the fine-tuned CLIP (ViT-L/14) significantly outperforms its zero-shot version, with Wikipedia MRR@10 reaching 0.7728 and overall MRR@10 at 0.6993. This stark improvement highlights the necessity of in-domain supervised fine-tuning for robust cross-domain retrieval. Among domains, Wikipedia consistently yields higher scores, reflecting pre-trained MLLMs’ inherent bias toward textual semantics. Formats like Slides, with sparse and fragmented content, pose greater challenges, reinforcing prior findings that zero-shot MLLMs underperform in complex, multi-modal retrieval scenarios.

\subsection{Multi-modal Fusion and Loss Exploration}

\paragraph{Weighted-sum fusion and BCE loss optimize cross-domain robustness} 
As shown in Table~\ref{tab:fusion_loss_comparison}, weighted-sum fusion with binary cross-entropy (BCE) loss achieves the best overall performance (MRR@10: 0.6566), significantly outperforming MLP fusion (0.2331). Introducing learnable MLP layers does not improve modality integration and may instead add noise or overfitting. BCE loss shares similarities with SigLIP’s sigmoid-based objective~\cite{siglip}, avoiding the large batch requirements of contrastive losses like InfoNCE~\cite{clip}, making it more efficient under limited computation.

\paragraph{BCE mitigates small-batch training challenges} 
Unlike InfoNCE, which depends on large batches to sample diverse negatives, BCE performs well with moderate batch sizes through per-example gradient updates. We further apply positive weighting in BCE to address class imbalance, where correct query-doc pairs are sparse. This adjustment increases model sensitivity to rare positives, following classic strategies in imbalanced learning~\cite{posweight1,posweight2}. Overall, BCE aligns better with the task’s data structure and hardware constraints.

\paragraph{Domain-specific modality demands highlight fusion simplicity} 
Domain-wise results support the robustness of simple fusion strategies: Wikipedia, being text-rich, benefits most from weighted-sum fusion (MRR@10: 0.7570), while Slides remains the most challenging due to sparse text and complex visuals (MRR@10: 0.3094). Across all settings, MLP-based fusion underperforms, reinforcing that in cross-domain retrieval, simplicity and generalisability outweigh the potential benefits of complex architectures.

\begin{figure*}[!h]
  \centering
  \begin{subfigure}[b]{0.48\linewidth}
    \centering
    \includegraphics[width=0.9\linewidth]{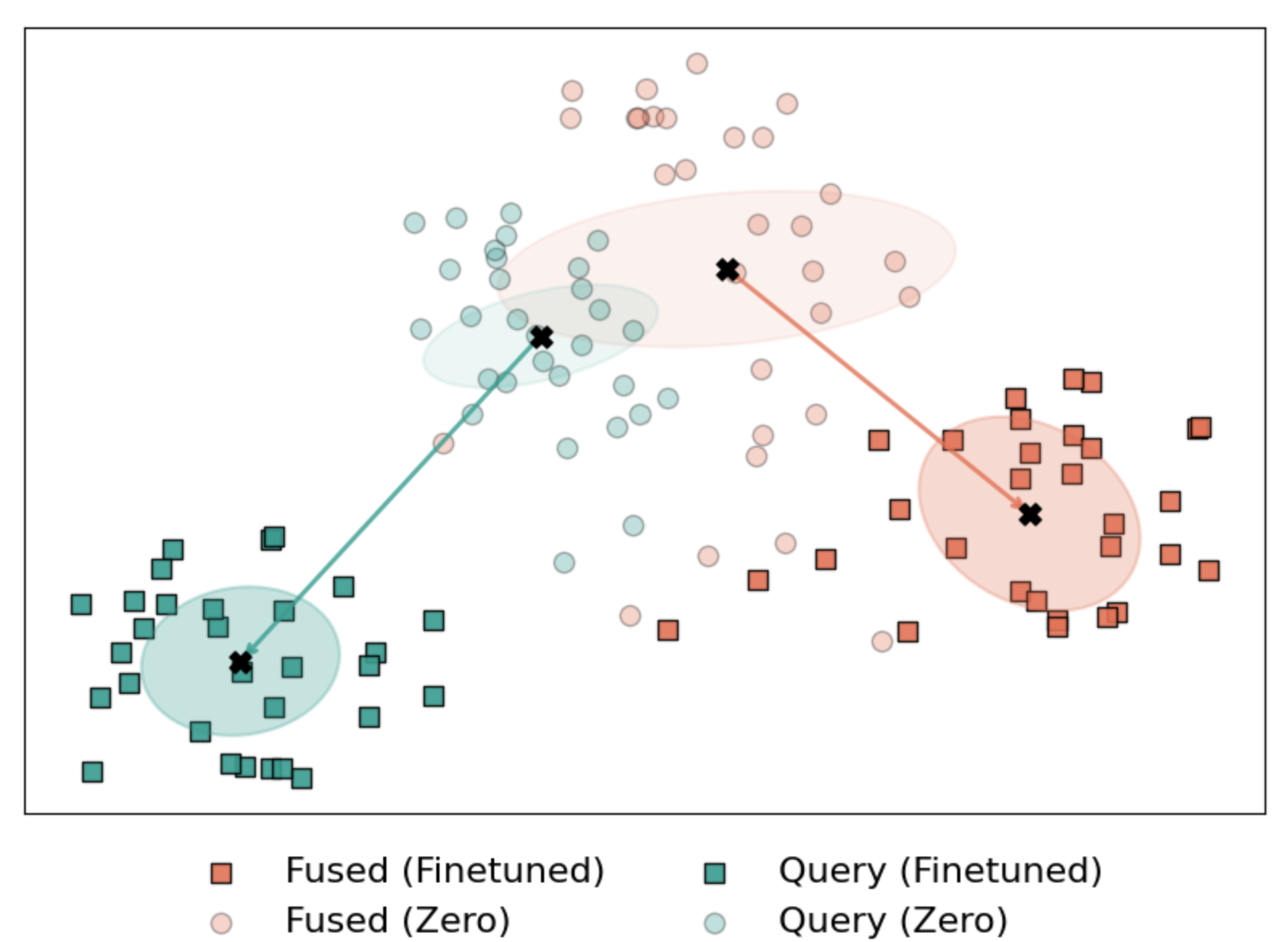}
    \caption{Query vs Text-Image fused embeddings before and after fine-tuning.}
    \label{fig:modality_shift}
  \end{subfigure}
  \hfill
  \begin{subfigure}[b]{0.48\linewidth}
    \centering
    \includegraphics[width=0.9\linewidth]{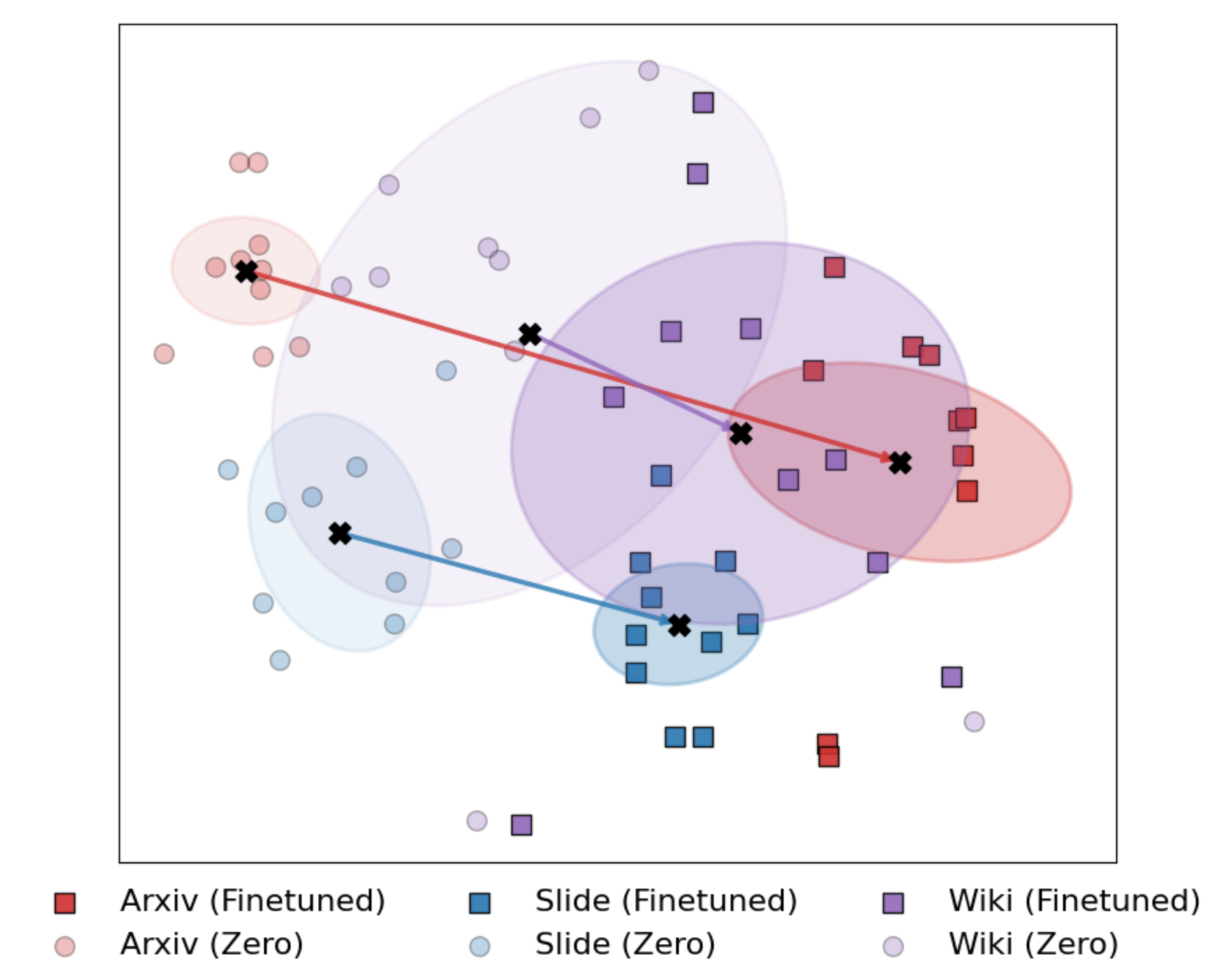}
    \caption{Query-Text-Image embedding shift across domains before and after training.}
    \label{fig:domain_shift}
  \end{subfigure}  
  \caption{t-SNE visualization of semantic embedding shifts before and after fine-tuning. 
  Shapes indicate training stage (circle = zero-shot, square = finetuned), ellipses denote cluster variance, and arrows indicate the shift of mean embeddings. }\label{fig:tsne-semantic-shift}
  \vspace{-4mm}
\end{figure*}

\subsection{Scaling and Training-Domain Impact}
\paragraph{Wikipedia exhibits superior cross-domain generalization.}
As demonstrated in \autoref{tab:training_scale_hit10}, models trained on Wikipedia achieve the strongest cross-domain performance, with a weighted average MRR@10 of \textbf{0.8758} at full data scale. This significantly outperforms arXiv-trained (\textbf{+16.6\%}) and Slides-trained (\textbf{+21.3\%}) models, underscoring Wikipedia’s text-rich structure as a catalyst for generalizable feature learning. 
The domain’s coherent textual semantics and broad topical coverage potentially enable models to transfer patterns effectively to other domains. 
In contrast, arXiv-trained models excel in their native domain (MRR@10: \textbf{0.7644}) but struggle with Slides and Wikipedia due to over-specialization in technical jargon and figures, while Slides-trained models suffer from sparse text and layout dependencies, achieving the lowest cross-domain performance.

\paragraph{Scaling reveals trade-offs between specialization and generalization}
Increasing training data within a single domain enhances in-domain performance but often compromises cross-domain adaptability. For arXiv, scaling from 20\% to 100\% data improves in-domain MRR@10 by \textbf{5.6\%} from 0.7236 to 0.7644, but reduces Slide set performance by \textbf{2.2\%} from 0.5810 to 0.5591. Reflecting a prioritization of domain-specific features. Similarly, Slides-trained models exhibit a \textbf{7.9\%} decline in cross-domain weighted average as training data scales, indicating amplified bias toward visual-centric patterns. Wikipedia-trained models show minimal degradation in cross-domain performance, suggesting near-saturation in learning transferable representations, though their inherent textual richness buffers against over-fitting.

\begin{table*}[htb]
\small
\centering
\resizebox{\linewidth}{!}{%
\begin{tabular}{p{3.2cm} p{3.3cm} p{2cm} p{5.5cm}}
\toprule
\textbf{Stage} & \textbf{Hardware \& Setting} & \textbf{Wall-time} & \textbf{Notes} \\
\midrule
Fine-tuning CLIP (360k train pairs, 5 epochs) 
& 8 $\times$ A100-80GB 
& $\approx$ 7.5 h 
& Total compute $\approx$ 60 GPU-h (parallelised). \\
\addlinespace[0.2em]
Embedding cache (390k page-windows, one-off) 
& 8 $\times$ A100-80GB 
& $\approx$ 30 min 
& Encodes the entire document collection once. \\
\addlinespace[0.2em]
Query encoding (15{,}000 test queries) 
& 1 $\times$ A100-80GB 
& $\approx$ 40 s 
& $\sim$2.7 ms per query. \\
\addlinespace[0.2em]
Retrieval with FAISS (IVF-HNSW) 
& 32-thread CPU 
& $\leq$ 0.5 s 
& $<$0.03 ms per query; negligible next to encoding. \\
\bottomrule
\end{tabular}}
\caption{Compute and memory analysis for the \textsc{Wiki} domain split.}
\label{tab:compute}

\end{table*}

\vspace{-4mm}
\subsection{Modality Ablation}

\paragraph{Text dominates retrieval signals, but visuals provide critical supplements}
The ablation study quantifies the impact of each modality. As illustrated in \autoref{tab:mrr10_modality_ablation}, Text-only models consistently outperform image-only ones, with full-set MRR@10 of 0.5333 vs. 0.1229. The gap is widest in Wikipedia (0.6286 vs. 0.1409), where dense text provides strong semantic cues, and narrowest in Slides (0.4661 vs. 0.3857), where sparse text limits retrieval quality and visuals add marginal value. These results confirm that lexical semantics remain the dominant retrieval signal, while images can still support performance in visually oriented domains.

\paragraph{Multi-modal synergy enhances cross-domain robustness}
Multi-modal models outperform unimodal ones overall (MRR@10: 0.6993 vs. 0.5333 for text-only), confirming the complementary value of image features. 
However, in Slides, fusion underperforms slightly (0.3223 vs. 0.3297), suggesting that naive combination of modalities may suppress sparse but critical textual cues. 
This aligns with the use of synthetic, text-based queries and highlights the difficulty of aligning fragmented visuals with lexical semantics. 
In contrast, arXiv (0.7144 vs. 0.2414) and Wikipedia (0.7728 vs. 0.6286) show clear multi-modal gains. 
These findings suggest future work should explore layout-aware or attention-based fusion to better integrate modalities in visually fragmented formats. 

\subsection{Visualization of Representation Shift}

To gain further insight into how the model learns and refines semantic representations, we visualize the embedding distributions before and after fine-tuning using t-SNE\footnote{\textbf{t-SNE parameters:} \texttt{n\_components=2}, \texttt{perplexity=30}, \texttt{n\_iter=1000}, \texttt{learning\_rate='auto'}, \texttt{random\_state=42}.}. Analysis focuses on two aspects: cross-modal alignment (query vs text-image fused) and domain-specific structure (doc-image).

\paragraph{Cross-modal.} \autoref{fig:modality_shift} shows the spatial layout of query and fused text-image embeddings. Before training, the two modalities are scattered and semantically distant. After training, both clusters become tighter and closer, with clear semantic migration indicated by arrows. This demonstrates that fine-tuning not only enhances intra-modal consistency but also improves cross-modal alignment, which is essential for retrieval performance.

\paragraph{Cross-domain.}
\autoref{fig:domain_shift} illustrates the distribution of query-text-image embeddings across three domains: Wikipedia, Slides, and arXiv. For Wikipedia, embeddings are already compact before training and remain stable after. ArXiv shows the most significant shift and convergence, reflecting increased semantic coherence. In contrast, the Slide domain exhibits the least improvement, likely due to sparse and fragmented text content, which makes visual-text fusion more difficult.

\subsection{Compute \& Memory Analysis}
\label{sec:compute}

To clarify feasibility, we report the compute and memory requirements on the \textsc{Wiki} domain split. 
Table~\ref{tab:compute} summarizes the wall-time and hardware settings for each stage of the pipeline. 
The results show that DocMMIR’s pipeline is practical at scale even on modest hardware, with the main cost being a 
one-off embedding cache rather than per-query computation.

\vspace{-1mm}
\section{Related Work}

\vspace{-2mm}
\paragraph{General Information Retrieval.}
Multi-modal benchmarks \cite{lin_2014_microsoft, young-etal-2014-image} have laid the groundwork for contrastive vision–language pretraining like CLIP~\cite{clip, align, siglip}, and later architectures such as BLIP-2 \cite{blip2} demonstrate parameter-efficient fusion of frozen image encoders with language models. Building on this, instruction-tuned and collaborative methods like UniIR \cite{uniir} and ColPali \cite{colpali} extend retrieval across multiple benchmarks, though they remain focused on instance-level matching rather than structured documents.

Concurrently, MM-Embed \cite{mmbed} and GME \cite{gme} leverage MLLMs to build universal multi-modal retrievers with strong pair-level alignment and instruction-following ability; however, they still operate on instance/chunked inputs and do not address document-level, mixed-granularity reasoning across pages.
\vspace{-2mm}
\paragraph{Multi-modal Document Retrieval.}
In document-centric settings, SciMMIR provides over 500K scientific figures and caption pairs for zero-shot and fine-tuned evaluation \cite{scimmir}, while MMDocIR \cite{mmdocir} introduces page-level and layout-level retrieval tasks but on limited data. 
The DSE paradigm \cite{dse} encodes entire document screenshots with MLLMs, excelling on Wikipedia and slide collections, and DocCVQA \cite{doccvqa} casts VQA as retrieval over 14,362 scanned pages. Despite their strengths, each targets specific domains or formats. DocMMIR addresses this by unifying diverse document types and retrieval scenarios.

\vspace{-3mm}
\section{Conclusion}
DocMMIR proposes a simple yet effective framework for unifying diverse document formats in cross-domain multi-modal retrieval. 
Experiments show that weighted-sum fusion with BCE loss enhances robustness across domains, while multi-domain training balances generalization and specialization—Wikipedia-trained models showing superior adaptability. Findings reveal that future work could explore layout-aware dynamic fusion, realistic query construction, and domain-balanced training to further improve on the DocMMIR task. 

\section*{limitations}

The DocMMIR framework has several limitations. Late fusion methods may be less effective for layout-heavy documents like Slides. While synthetic queries offer scalability and control, future work could explore complementing them with real user queries to enhance realism. The domain imbalance (e.g., 79.8\% Wikipedia) reflects practical differences in data accessibility across domains. Lastly, some recent retrieval baselines (e.g., VLM2Vec, DSE) are not included due to limited public implementations.

\section*{Ethics Statement}
The development and deployment of the DocMMIR framework raises several ethical considerations. First, the datasets utilized in this work: Wikipedia articles, arXiv papers, and Slideshare presentations, are publicly available, but their use necessitates careful adherence to licensing and attribution requirements. For Slideshare data, user-generated content may include sensitive or proprietary material; we mitigated this risk by employing OCR to extract only publicly visible text and filtering out documents with potential privacy violations (e.g., personal identifiers). Second, query generation via large language models (LLMs) introduces potential biases inherent in the base model (Qwen2.5-VL). While we manually validated query relevance to minimize hallucination or misalignment, the synthetic nature of queries may not fully represent real-world user intent, particularly in culturally or technically nuanced contexts.

\bibliography{custom}
\bibliographystyle{acl_natbib}

\appendix

\section{Domain-specific Multi-modal Characteristics}

To better understand the diversity and multi-modal nature of our dataset, we conduct a qualitative analysis by sampling 20 representative documents from each domain. Below, we describe the distinct characteristics observed across the three major sources:

\subsection{arXiv}
Documents from \textit{arXiv} are typically research papers from disciplines such as physics, computer science, and mathematics. They contain technical plots, detailed diagrams, and dense textual content with frequent use of mathematical formulas and citations. Visual elements (e.g., charts, tables, algorithm pseudocode) are tightly integrated with the surrounding text to support analytical reasoning. The corresponding user queries often begin with ``How'' or ``Why'', reflecting a desire for deep technical understanding and methodological insights.

\subsection{Slideshare}
\textit{Slideshare} content consists primarily of presentation slides, which are highly structured with section headings, bullet points, and concise text. Visuals include simplified illustrations, conceptual diagrams, and occasional infographics. The language is generally less technical and more pedagogical, aiming for clarity and accessibility. Topics span education, business, media, and more. Queries typically focus on understanding core concepts or practical applications, aligning with the explanatory intent of the documents.

\subsection{Wikipedia}
\textit{Wikipedia} articles follow an encyclopedic style, with informative images such as historical photographs, maps, and diagrams that supplement rich textual descriptions. These documents are entity-dense and often include timelines, factual summaries, and citation references. The multi-modal elements are designed to support broad-topic comprehension. User queries associated with Wikipedia data frequently explore cultural, historical, or factual context, emphasizing general knowledge acquisition.

\section{Evaluation Protocol and Metric Definitions}
\label{app:eval}

Our evaluation adopts a \emph{full-collection} retrieval paradigm: every test query is ranked against the entire document pool, including training, validation, and test items.  This mirrors deployment settings in which retrieval engines must sift relevant documents from a continually growing corpus.  The design (i) prevents inflated scores that arise when ranking is restricted to unseen documents, (ii) reflects practical scenarios lacking strict train–test boundaries, and (iii) discourages shortcut learning by forcing the model to discriminate relevant from irrelevant material regardless of split origin.

\paragraph{MRR@10.}  For each query, we compute the reciprocal rank of the first relevant document within the top-10 results and average across queries.  By truncating at ten, MRR@10 focuses evaluation on the page of results most users inspect.

\paragraph{NDCG@10.}  Normalised Discounted Cumulative Gain at rank ten measures graded relevance, weighting higher-ranked documents more heavily and normalising by the ideal ranking.

\paragraph{HIT@$K$.}  HIT@1, HIT@3 and HIT@10 report the proportion of queries for which at least one relevant document appears within the top $K$ positions, providing an intuitive precision-style view.

Together, these complementary metrics yield a robust assessment of ranking quality in large candidate pools while emphasising practical user-facing cut-offs.

\section{Additional Experimental Results}
\label{app:colpali}

\subsection{ColPali under DocMMIR}
We also evaluated ColPali on our benchmark to examine whether models trained for page-screenshot retrieval generalize to our document-level setting. 
Table~\ref{tab:colpali} reports zero-shot results across three domains.

\begin{table}[ht]
\centering
\resizebox{\linewidth}{!}{%
\begin{tabular}{lccc}
\toprule
\textbf{Domain} & \textbf{Both Modalities (ZS)} & \textbf{Text-only (ZS)} & \textbf{Image-only (ZS)} \\
\midrule
arXiv & 0.0002 & 0.0002 & 0.0002 \\
Slide & 0.0010 & 0.0008 & 0.0008 \\
Wiki  & 0.0000 & 0.0000 & 0.0000 \\
\bottomrule
\end{tabular}}
\caption{Zero-shot performance on \textbf{MRR@10} of ColPali on the \textit{wiki} domain.}
\label{tab:colpali}
\end{table}

As shown in Table~\ref{tab:colpali}, ColPali performs extremely poorly in our benchmark. 
The main reason is a \textbf{setting mismatch}: ColPali is trained for \emph{query~$\leftrightarrow$~page-screenshot} retrieval, whereas our task requires fine-grained reasoning over text--image units within \emph{multi-page documents}. 
This discrepancy explains the near-zero scores. Importantly, ColPali’s architecture is not inherently weak, but it is misaligned with a task that demands dynamic granularity and cross-domain supervision---precisely the gap that our DocMMIR pipeline is designed to fill. 

For this reason, we did not extend evaluation to other recent MLLMs that share the same screenshot-based retrieval setting (e.g., DSE-style models): their training objectives and input assumptions similarly diverge from DocMMIR’s document-level, mixed-granularity requirements, and thus their reported performance would not provide additional insight beyond what we observe with ColPali.

\subsection{Fine-tuned Baseline Models}
We also conducted preliminary fine-tuning experiments on the \textit{arXiv} domain to examine whether other MLLMs can close the gap with simple domain adaptation. 
Table~\ref{tab:finetuned} summarizes these results.

\begin{table}[h]
\centering
\small
\begin{tabular}{lcc}
\toprule
\textbf{Fine-tuned Models} & \textbf{MRR@10} & \textbf{Hit@10} \\
\midrule
SigLIP2 (base-patch16-224) & 0.08932 & 0.17952 \\
SigLIP2 (so400m-patch14-384) & 0.14046 & 0.24612 \\
ALIGN & 0.33590 & 0.61735 \\
BLIP-2 & 0.00008 & 0.00033 \\
BLIP-2 + BERT & 0.00033 & 0.00033 \\
CLIP + BERT & 0.00015 & 0.00066 \\
\bottomrule
\end{tabular}
\caption{Fine-tuned performance on \textbf{MRR@10} of several baseline models on the \textit{arXiv} domain.}
\label{tab:finetuned}
\end{table}

The results show that fine-tuning alone cannot overcome the cross-modal difficulties posed by our task. 
Pre-trained MLLMs such as CLIP, ALIGN, SigLIP-2, and BLIP-2 mainly learn global image--sentence correspondences. 
After fine-tuning, some (e.g., SigLIP-2, ALIGN) improve modestly but still miss many correct matches and plateau at relatively low MRR@10 / Hit@10. 

We further evaluated two ``vision+text'' variants: BLIP-2 + BERT-text and CLIP-vision + BERT-text. Both collapse to almost zero performance (Hit@10 $\leq$ 0.00066). This failure is due to three interconnected reasons:  
(i) \textbf{lack of joint pre-training}: vanilla BERT has never been exposed to visual embeddings, making cross-modal alignment from scratch infeasible;  
(ii) \textbf{distribution mismatch}: the statistical properties of BERT’s token embeddings differ markedly from vision encoder patch features, leading to unstable gradients;  
(iii) \textbf{wrong granularity}: BERT captures sentence-/token-level semantics, whereas DocMMIR requires region- and layout-level cues.  

Taken together, these experiments demonstrate that simply plugging in a ``stronger'' text encoder is insufficient. The core bottleneck is robust cross-modal ability, not richer text semantics.

\section{Dataset Example}
\label{app:example}

\begin{figure*}[h]
\centering
\includegraphics[width=1\textwidth]{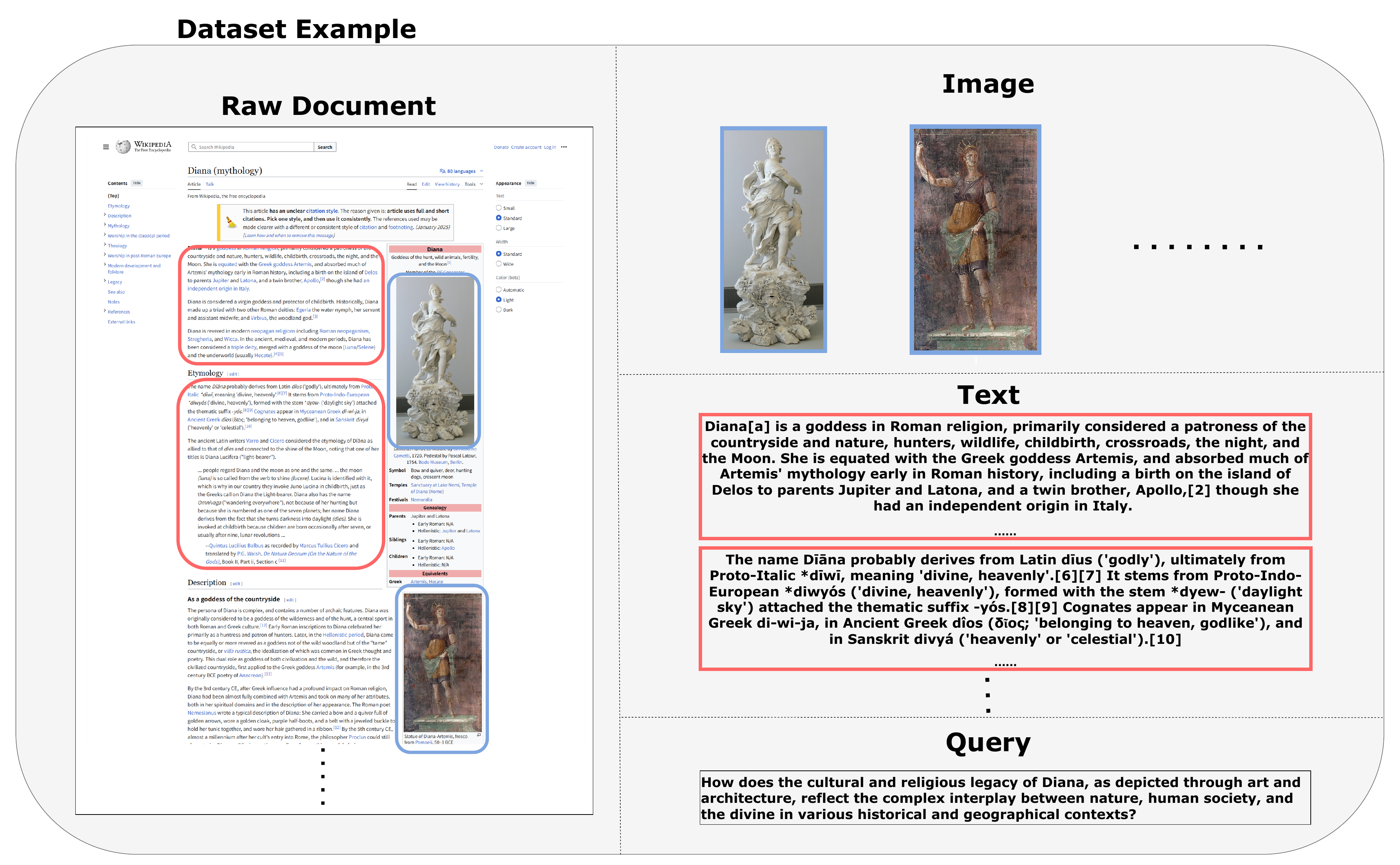}
\caption{Example entry from the DocMMIR dataset, showing a document excerpt, texts, linked images, and associated query. 
Source document: Diana (mythology) (Wikimedia Commons).}
\label{fig:data}
\vspace{-4mm}
\end{figure*}

To illustrate the structure of a typical DocMMIR entry, we provide one representative example below in \autoref{fig:data}. 
Each entry integrates a set of document texts, a set of associated images, and a text query. 

\section{Prompt Template}
\label{app:prompt}

Prompts used in the data curation pipeline are listed below.

\begin{promptbox}[box:query_prompt]{Prompt for Query Generation}
\begin{verbatim}

{ "role": "system",
  "content": "You are a helpful natural language processing expert." },
{ "role": "user",
"content": [
    *doc_image,
    {
        "type": "text",
        "text": (
            "You are tasked with generating a thought-provoking question "
            "based on the given image-text data from a document. "
            "The question should capture the overall theme or deeper "
            "meaning of the document, rather than specific visual details. "
            "It must be abstract, invite critical reflection, and avoid "
            "a direct answer from the context. "
            "Do not be overly generic—ensure the question aligns with "
            "the unique visual cues of the document. "
            "Begin your output with 'Q:' followed by the generated question. "
            + doc_text
        )
    }
  ]
}

\end{verbatim}
\end{promptbox}

\begin{promptbox}[box:ocr]{Prompt for Slide OCR data perplexity examination}
\begin{verbatim}
{
    "role": "system",
    "content": (
        "You are a generous language quality classifier. "
        "Your task is to determine whether a given text segment, possibly "
        "extracted from an OCR-processed document, likely contains meaningful "
        "human-written content. You should accept text that is partially "
        "broken, informal, or noisy, as long as it seems intended to "
        "communicate something relevant. Accept marketing language, product "
        "descriptions, announcements, or technical explanations. Only reject "
        "text if it is purely noise, random symbols, or unreadable junk. "
        "/no_think"
    )
},
{
    "role": "user",
    "content": (
        "Below are some examples:\n\n"
        "Text: 'Figure 3: 0.233!!@@## 19982ab' → No\n"
        "Text: 'Explori enables survey management for licensed events.' → Yes\n"
        "Text: 'Chart axis: year, value, growth' → No\n"
        "Text: 'This document introduces a framework for multi-modal IR tasks "
        "in scientific domains.' → Yes\n"
        "Text: 'http://bit.ly/xyz download summary' → No\n"
        "Text: 'Project overview and next steps: iterate, test, deploy' → Yes\n"
        "---\n\n"
        "Now classify the following:\n\n"
        f"Text: {doc_text}\n\n"
        "Is this meaningful human language? Respond with one word only: "
        "'Yes' or 'No'."
    )
}
\end{verbatim}
\end{promptbox}

\end{document}